\documentclass[10pt,twocolumn,letterpaper]{article}

\usepackage{cvpr} 
\usepackage{times}
\usepackage{epsfig}
\usepackage{graphicx}
\usepackage{amsmath}
\usepackage{amssymb}

\usepackage[pagebackref=true,breaklinks=true,letterpaper=true,colorlinks,bookmarks=false]{hyperref}

\begin{document}

\title{Beyond EER: Multi-Dimensional Evaluation of Information Leakage\\in Speaker De-Identification}

\author{Seungmin Seo\\
Contractor Associate, National Institute of Standards and Technology, Gaithersburg, MD, USA\\
Chakra Consulting Inc., Clarksburg, MD, USA\\
\and
Oleg Aulov\quad
P. Jonathon Phillips\quad
Kevin Mangold\quad
Jonathan Eskin\\
National Institute of Standards and Technology, Gaithersburg, MD, USA\\
{\tt\small \{seungmin.seo, oleg.aulov, jonathon.phillips, kevin.mangold, jonathan.eskin\}@nist.gov}
}

\maketitle
\thispagestyle{empty}

\begin{abstract}
Speaker de-identification (SDID) aims to preserve privacy by concealing speaker identity while maintaining speech utility. However, current evaluations often reduce privacy to a single dimension—biometric verification performance—typically measured by Equal Error Rate (EER). This narrow focus ignores critical leakage channels, such as soft biometric inference, embedding-level re-identification, and structural template similarity, which threaten the unlinkability and irreversibility of biometric references. We propose a holistic evaluation framework across five complementary metrics: (i) EER, (ii) soft biometric leakage score , (iii) cumulative match characteristic re-identification analysis, (iv) canonical correlation analysis and Procrustes embedding alignment, and (v) intelligibility via word error rate and semantic similarity. Evaluating five SDID systems from the IARPA ARTS program, we demonstrate that these metrics capture independent dimensions of information leakage. Our results indicate that reliance on a single metric can misrepresent the privacy properties of an SDID system.
\end{abstract}

\section{Introduction}
Speech signals encode biometric and behavioral attributes beyond linguistic content, implicitly revealing a speaker's sex, age, and accent~\cite{Casandra2023IJCB, lin2026toward}. Speaker de-identification (SDID) systems aim to transform these signals to remain intelligible while concealing the speaker's identity~\cite{tomashenko2022voiceprivacy,panariello2024voiceprivacy, tomashenko2024voiceprivacy, Li2025HLTCOEAttacker}. However, the growing availability of high-accuracy, publicly accessible attribute classifiers~\cite{yang2025demographic} has fundamentally changed the threat landscape: modern speech representation models enable accurate inference of biometric attributes even from anonymized speech. This raises a central question: \textbf{How much identity information remains detectable in speech processed by current SDID systems, and along which dimensions can it be measured?}

This pattern echoes findings in other biometric modalities: in face recognition, embeddings have been shown to inadvertently store demographic, paralinguistic, and extrinsic factors beyond identity~\cite{PardeFG17,Hill2018DeepCN, dhar2021pass,phillips2025state}. Yet while the VoicePrivacy Challenges~\cite{tomashenko2022voiceprivacy, tomashenko2024voiceprivacy, panariello2024voiceprivacy, tomashenko2024first} have established a benchmark for voice anonymization, evaluation of these residual signals in speech remains fragmented.

Most current evaluations rely on a solitary metric such as Equal Error Rate (EER) for speaker verification~\cite{tomashenko2022voiceprivacy,panariello2024voiceprivacy, Li2025SpecWavAttack,Lyu2025FastAdaptation,Mawalim2025TitaNetAttacker,Zhang2025VoiceAttack, tomashenko2024voiceprivacy}, which captures resistance to direct re-identification but not whether an adversary can recover demographic attributes or exploit systematic patterns across a population. Privacy metrics alone are also insufficient without understanding their cost in utility: a system that renders speech unintelligible achieves perfect privacy at the expense of speech communication's fundamental purpose.

This paper makes the following contributions{\let\thefootnote\relax\footnote{Disclaimer: Certain equipment, instruments, software, or materials are identified in this paper in order to specify the experimental procedure adequately. Such identification is not intended to imply recommendation or endorsement of any product or service by NIST, nor is it intended to imply that the materials or equipment identified are necessarily the best available for the purpose. These opinions, recommendations, findings, and conclusions do not necessarily reflect the views or policies of NIST or the United States Government.}}\setcounter{footnote}{0}:

\begin{enumerate}
\item \textbf{Evaluation across complementary privacy metrics.}
We evaluate SDID systems using various privacy metrics spanning verification resistance (EER), attribute leakage (SBLS), search-based re-identification using cumulative match characteristic (CMC), and embedding-space similarity measured by canonical correlation analysis (CCA) and Procrustes alignment. These metrics capture distinct aspects of residual speaker information and can yield different assessments of privacy.

\item \textbf{Biometric leakage analysis on anonymized speech.}
We extend SBLS\cite{Seo2025SoftBiometricLeakage} to include accent prediction and introduce subgroup protection analysis to measure demographic disparities in attribute leakage.

\item \textbf{Large-scale cross-corpus evaluation.}
We evaluate SDID systems across four test sets from three corpora (Mixer 3, 6, and 7), covering multiple accents and sampling conditions with more than 22,000 segments per system and over 3.4 million verification trials.

\item \textbf{Privacy-utility quantification.}
We measure the impact of anonymization on intelligibility using word error rate (WER) and semantic similarity, revealing a measurable trade-off between identity suppression and speech utility.

\end{enumerate}

Comparing system performance across these dimensions, we find that no system achieves top performance across all of them: the metrics capture fundamentally different aspects of privacy that can trade off against one another, and reliance on any single metric yields an incomplete—and potentially misleading—assessment.

\section{Related Work}

\subsection{Speaker De-Identification Systems and Evaluation}

SDID spans signal-processing transforms and neural voice conversion. The VoicePrivacy Challenge~\cite{tomashenko2022voiceprivacy, panariello2024voiceprivacy, tomashenko2024voiceprivacy, tomashenko2024first} established a common benchmark, primarily evaluating systems using EER and log-likelihood ratio cost. Subsequent work explores diverse approaches, including speaker attribute perturbation~\cite{11095803}, voice conversion with alternative distance metrics and kNN conversion in self-supervised spaces~\cite{Li2025HLTCOEAttacker}. Despite architectural diversity, evaluation remains largely centered on speaker verification, offering limited insight into demographic leakage or privacy--utility trade-offs.

\subsection{Identity Information Leakage in Anonymized Speech}

Recent work shows that anonymized speech can still contain residual identity information detectable through modern speaker representation models~\cite{Li2025SpecWavAttack,Mawalim2025TitaNetAttacker,Zhang2025VoiceAttack,Tomashenko2025TemporalDynamics,Lyu2025FastAdaptation}. These studies primarily evaluate speaker-level resistance using speaker verification metrics.

Beyond speaker identity, speech embeddings also encode soft biometric attributes such as sex, age, and accent~\cite{yang2025demographic}. VoxProfile~\cite{Feng2025VoxProfile} demonstrates large-scale extraction of speaker traits from foundation-model embeddings, while prior studies on benchmark dataset usage dynamics~\cite{Casandra2023IJCB} highlight privacy challenges.

Together, these findings suggest that resistance to speaker re-identification alone does not guarantee broader privacy protection. Anonymized speech may still reveal demographic attributes or structural embedding similarities. This motivates evaluating SDID systems across multiple complementary metrics of privacy and utility.

\section{Evaluation Setup}

\subsection{Corpora and Test Conditions}

Evaluation data were developed by NIST from selected segments of the Mixer 3, 6, and 7 corpora collected by the Linguistic Data Consortium (LDC). Segments of 10, 30, and 60 seconds were generated using the LDC Broad Phonetic Class Speech Activity Detector~\cite{ldc-bpcsad}. The first 60 seconds of each recording were discarded to exclude greetings and channel stabilization effects. Test and initialization recordings were disjoint.

\begin{table*}[t]
\centering
\begin{tabular}{lccccc}
\hline
\textbf{} & \textbf{Test 1 (16k)} & \textbf{Test 1 (8k)} & \textbf{Test 2 (8k)} & \textbf{Test 3 (16k)} & \textbf{Test 4 (8k)} \\
\hline
Original corpus & Mixer 6 & Mixer 6 & Mixer 3 & Mixer 7 & Mixer 3 \\
Total trials & 277,634 & 277,634 & 951,025 & 991,474 & 1,244,014 \\
Target trials & 4,838 & 4,838 & 10,458 & 52,300 & 77,434 \\
Non-target trials & 272,796 & 272,796 & 940,567 & 939,174 & 1,166,580 \\
Unique speakers & 76 & 76 & 223 & 60 & 76 \\
Unique segments & 1,058 & 1,058 & 2,983 & 2,778 & 3,355 \\
Male speakers & 43 & 43 & 81 & 13 & 37 \\
Female speakers & 33 & 33 & 142 & 47 & 39 \\
Adult speakers (25--54) & 47 & 47 & 161 & 40 & 63 \\
Senior speakers (55--85) & 4 & 4 & 30 & 20 & 13 \\
Young speakers (18--24) & 25 & 25 & 32 & -- & -- \\
Native language & English & English & English & Spanish & Hindi \\
\hline
\end{tabular}
\caption{Statistics of evaluation dataset.}
\label{tab:eval_dataset}
\end{table*}

The test sets were designed to address complementary evaluation goals. Test 1 enables controlled sample-rate comparison (8~kHz vs.\ 16~kHz) on the same speakers using Mixer 6 CHiME interview recordings with available human transcriptions. Test 2 uses Mixer 3 8 kHz sampled English data. Test 3 and 4 introduce linguistic diversity: Mixer 7 Spanish speakers and Mixer 3 Hindi speakers producing English speech, enabling analysis of non-native accent handling. The statistics of test sets are shown in Table~\ref{tab:eval_dataset}.

The Mixer/ARTS corpora were selected over alternatives such as the VoicePrivacy Challenge data because they provide the necessary speaker metadata, including annotations for sex, age group, and native language that are not available in existing privacy-evaluation benchmarks.

Demographic distributions are imbalanced across test sets. Test 3 contains no Young speakers and is 73\% Female; Test 4 is sex-balanced but skewed Adult; These conditions reflect realistic deployment scenarios and expose potential subgroup vulnerabilities.

\subsection{Trials}
We employ five trial types: three designed to assess resistance to speaker identification (SID) attacks, one to evaluate pseudo-identity consistency, and one to measure pseudo-identity distinctness. Here, a pseudo-identity refers to the anonymized speaker profile produced by the SDID system, where each original speaker is mapped to a consistent synthetic identity across utterances (e.g., Alice → pseudo-Alice). Table~\ref{tab:trial} summarizes the trial types.

\begin{table*}[t]
\centering
\resizebox{0.80\linewidth}{!}{%
\small
\begin{tabular}{llcl}
\hline
\textbf{Trial type} & \textbf{Composition (target vs. non-target)} & \textbf{Target EER} & \textbf{Purpose} \\
\hline
\textit{oaoa} & orig-anon vs.\ orig-anon & 50\% & SID attack resistance \\
\textit{oaoo} & orig-anon vs.\ orig-orig & 50\% & SID attack resistance \\
\textit{oaaa} & orig-anon vs.\ anon-anon & 50\% & SID attack resistance \\
\textit{aaaa} & anon-anon vs.\ anon-anon & 0\% & Self-consistency \\
\textit{cross-profile} & p1-p1 vs.\ p1-p2 (cross pseudo-profile) & 0\% & Pseudo-identity distinctness \\
\hline
\end{tabular}
}
\caption{Five different trial types used in evaluation.}
\label{tab:trial}
\end{table*}

A well-performing anonymization system should produce voices that cannot be traced back to the original speaker. Pseudo-speaker output should functionally behave like real speaker output in that SID systems consistently match the same pseudo-speaker to itself, and not to other pseudo-speakers or real speakers.

The three SID attack trials assess different aspects of this goal. In \textit{oaoa}, original segments are compared against anonymized segments (e.g., Alice vs.\ pseudo-Bob), directly testing whether an attacker can match a known voice to its anonymized counterpart. In \textit{oaoo}, original segments are compared against other original recordings (e.g., Alice vs.\ Bob), testing whether the presence of anonymized enrollment data degrades the SID backend's ability to distinguish real speakers. In \textit{oaaa}, anonymized segments are compared against other anonymized segments (e.g., pseudo-Alice vs.\ pseudo-Bob), testing whether different anonymized speakers remain distinguishable from one another. EER near 50\% indicates strong performance on each. However, an inflated \textit{oaaa} EER (well above 50\%) can indicate that anonymized voices from different speakers have become acoustically similar, compressing the speaker identity space rather than successfully concealing individual identities.

The \textit{aaaa} trial evaluates self-consistency with a target EER of 0\%. Pseudo-Alice should consistently sound like pseudo-Alice, regardless of which utterance is anonymized. Higher EER indicates that different segments from the same speaker may sound like different people after processing. The \textit{cross-profile} trial evaluates pseudo-identity distinctness, also targeting 0\% EER. Pseudo-Alice should be clearly distinguishable from pseudo-Bob; an EER near 50\% would indicate that different pseudo-identities are acoustically interchangeable.

\subsection{Speaker De-Identification Systems}

Five SDID systems developed under the IARPA ARTS\footnote{https://www.iarpa.gov/research-programs/arts} program were evaluated: four performer submissions and one baseline system. The systems span diverse architectural approaches:

\textbf{VOXLET} maps audio to wav2vec 2.0 latent representations, applies differential privacy noise in latent space, and reconstructs speech using HiFiGAN 2.0.

\textbf{RASP} employs a disentangled autoencoder separating content (HuBERT), speaker identity, and pitch/energy. Speaker identity is replaced with a pseudo-speaker embedding selected via cosine similarity.

\textbf{SHADOW} uses an autoregressive language model over EnCodec tokens conditioned on Wav2Vec2 features. Pseudo-speaker embeddings are generated via PLDA and converted with FreeVC.

\textbf{PHORTRESS} decomposes speech using the SPARC articulatory coding framework. The speaker embedding is replaced with a fabricated identity, and a DDSP vocoder resynthesizes speech from articulatory features.

\textbf{Baseline} system performs k-nearest-neighbor regression on WavLM features, averaging matched pseudo-speaker embeddings and synthesizing with HiFiGAN.

\subsection{Speaker Identification Backends}

Four independent SID backends assess anonymization effectiveness:

\begin{itemize}
\item \textbf{NeMo TitaNet Large:} depth-wise separable convolutions with SE layers and channel-attention pooling, trained on VoxCeleb 1\&2, Fisher, SwitchBoard, LibriSpeech, and augmented data~\cite{Koluguri2022TitaNet}.
\item \textbf{NeMo ECAPA-TDNN:} TDNN with SE and Res2Block layers and multi-scale attention~\cite{Dawalatabad2021ECAPA}.
\item \textbf{Hyperion:} ResNet-based x-vector extractor with PLDA backend trained on NIST SRE CTS Superset. \cite{villalba2022advances}
\item \textbf{OLIVE:} TDNN x-vector extractor with PNCC features and PLDA backend trained on NIST SRE 2004--2012, Mixer6, and VoxCeleb 1\&2 \cite{lawson2016open}.
\end{itemize}

\subsection{Attribute Classifier and Automated Speech Recognition Systems}

\textbf{VoxProfile}~\cite{Feng2025VoxProfile} serves as a strong attribute inference adversary for evaluating soft biometric leakage. While greater attacker diversity would offer a more complete assessment, using a single threshold-free attacker in our analysis reduces sensitivity to classifier calibration. Evaluated attributes include sex (binary), age group (Young 17--24, Adult 25--54, Senior 55+), and accent family (North America, Romance, South Asia). Accent is evaluated only on aggregated cross-test results.

For intelligibility assessment, OpenAI Whisper~\cite{Radford2023Whisper} and NVIDIA NeMo Canary-1B~\cite{Sekoyan2025Canary} transcribe original and anonymized speech. Canary-1B was additionally used because it exhibits fewer hallucinations during silent segments. The Whisper English text normalizer (lowercasing, contraction expansion, punctuation removal, numeric normalization) is applied for WER scoring.

\section{Evaluation Metrics}

We evaluate SDID systems along five complementary dimensions capturing speaker-level privacy using EER, soft biometric leakage using SBLS, re-identification risk using CMC, embedding structure using CCA, and utility preservation using WER.

\subsection{Speaker Verification EER}

EER measures the operating point at which the false acceptance rate equals the false rejection rate in a speaker verification task. EER results are aggregated across SID backends and pseudo-identity profiles. Higher EER in SID attack trials (target 50\%) indicates stronger privacy; lower EER in consistency trials (target 0\%) indicates reliable pseudo-identity behavior.

\subsection{Soft Biometric Leakage Score (SBLS)}

SBLS~\cite{Seo2025SoftBiometricLeakage} quantifies soft biometric leakage via three components:

\begin{equation}
\text{SBLS} = \alpha P_{\text{attr}} + \beta P_{\text{assoc}} + \gamma P_{\text{subgroup}},
\end{equation}

where $\alpha,\beta,\gamma \ge 0$, $\alpha+\beta+\gamma=1$, and each component lies in $[0,1]$ (1 = maximal privacy). In this work, we set $\alpha=\beta=0.4$ and $\gamma=0.2$ to slightly downweight subgroup robustness, though the choice is heuristic. The impact of parameter is reported in section~\ref{sec:weight_sensitivity}

\begin{enumerate}
\item \textbf{Zero-Shot Attribute Privacy ($P_{\text{attr}}$):}
Let $A$ be the set of attributes (e.g., $A=\{\text{Male/Female label},\allowbreak \text{age group}\}$). For each attribute $a\in A$ with $K_a$ classes, the frozen attacker outputs class scores on the anonymized dataset. We compute one-vs-rest AUC for each class from scores.

\begin{equation}
P_{\text{attr}} =
1 - \frac{1}{|A|} \sum_{a \in A}
\frac{\max\{0, \text{mAUC}_a^* - 0.5\}}{0.5}.
\end{equation}

Here $P_{\text{attr}}{=}1$ indicates chance-level zero-shot inference (low leakage) and $P_{\text{attr}}{=}0$ indicates near-perfect attribute recoverability (high leakage). When only hard predictions are available, we substitute macro balanced accuracy for $\mathrm{mAUC}_a^\star$ and normalize by the chance-level baseline $1/K_a$.

\item \textbf{Systematic Association ($P_{\text{assoc}}$):}
Let $\hat A_a^\star$ denote hard predictions obtained via $\arg\max$ over the permutation-aligned scores from the previous step.
Residual statistical dependence between true and predicted labels is measured using normalized mutual information:

\begin{equation}
P_{\text{assoc}} =
1 - \frac{1}{|A|} \sum_{a \in A}
\frac{I(A_a;\hat{A}_a^*)}{\log K_a}.
\end{equation}

where $I(A_a; \hat A_a^\star)$ is the mutual information between true and predicted attributes. In practice, we estimate $\tilde I_a$ from the confusion matrix between $(A_a, \hat A_a^\star)$ using standard entropy calculations. High $\tilde I_a$ indicates strong systematic dependence (higher leakage), while $\tilde I_a \approx 0$ suggests that predictions contain little information about true attributes beyond random chance.

\item \textbf{Subgroup Protection ($P_{\text{subgroup}}$):}
Equity across demographic subgroups is evaluated as

\begin{equation}
P_{\text{subgroup}} =
\omega (1 - \max_g L_g)
+
(1-\omega)
\frac{\min_g(1 - L_g)}{\max_g(1 - L_g)},
\end{equation}

where $L_g$ denotes subgroup leakage and $\omega=0.7$. Varying $\omega \in \{0.3, 0.5, 0.7, 0.9\}$ changes SBLS by at most 0.004 and does not affect rankings. When accent is included, demographic and accent subgroup protection are averaged equally. 

\end{enumerate}

\subsection{CMC}

CMC simulates a closed-set identification attack. For each anonymized embedding, cosine distances to all original embeddings are computed, and the rank of the first correct match is recorded:

\begin{equation}
\text{CMC}@k =
\frac{1}{N}
\left|
\{ i : \text{rank}_i \le k \}
\right|.
\end{equation}

Lower CMC values indicate stronger anonymization. Both NeMo embeddings are used. A permutation bootstrap establishes chance-level performance.

\subsection{CCA and Procrustes Alignment}

CCA measures linear relationships between original and anonymized embedding subspaces. CCA is fit on matched pairs (80/20 train/test split), and the mean of the top-10 canonical correlations is reported on held-out data.

Procrustes alignment learns an orthogonal rotation $R$ minimizing alignment error and evaluates alignment via mean cosine similarity.

A random permutation baseline breaks original–anonymized pairing, yielding reference levels (CCA Top-10 $\approx 0.734$, Procrustes cosine $\approx 0.29$--$0.31$). Values near baseline indicate decorrelation; values approaching 1 indicate linear predictability.

\subsection{Speech Intelligibility and Semantic Preservation}

Utility cost is measured using two ASR systems: OpenAI Whisper and NVIDIA NeMo Canary-1B. WER is computed as

\begin{equation}
\text{WER} = \frac{S + D + I}{N},
\end{equation}

where $S$, $D$, and $I$ denote substitutions, deletions, and insertions, and $N$ is the number of reference words.

Because WER penalizes all errors equally regardless of semantic impact, we complement it with cosine similarity of sentence embeddings~\cite{Song2020MPNet}. Cosine similarity ranges from 0 to 1, with higher values indicating stronger semantic preservation.

\begin{table}[t]
\centering
\small
\resizebox{0.98\linewidth}{!}{%
\begin{tabular}{lccccc}
\hline
\textbf{SID Backend} & \textbf{Test 1 (16k)} & \textbf{Test 1 (8k)} & \textbf{Test 2} & \textbf{Test 3} & \textbf{Test 4} \\
\hline
NeMo TitaNet & 4.75 & 29.70 & 3.82 & 4.51 & 6.49 \\
NeMo ECAPA & 4.81 & 22.14 & 3.64 & 5.21 & 5.06 \\
Hyperion & 4.70 & 7.55 & 4.86 & 5.50 & 4.48 \\
SRI OLIVE & 2.60 & 6.06 & 4.38 & 4.19 & 3.45 \\
\hline
\end{tabular}
}
\caption{Reference EER (\%) on original speech for each SID backend across different test dataset.}
\label{tab:orig_eer_by_backend}
\end{table}

\section{Evaluation Results}

\subsection{EER-Based Speaker Verification Results}
We first report EER-based evaluation to establish the speaker-level privacy landscape before examining soft biometric and embedding-structure dimensions.

Table~\ref{tab:orig_eer_by_backend} shows the reference performance on original speech with different SID backends. All four SID backends achieve EER below 7\% on most conditions, except Test 1 at 8~kHz, where TitaNet (29.7\%) and ECAPA-TDNN (22.1\%) show degraded performance. Hyperion (7.6\%) and OLIVE (6.1\%) remain robust.

\begin{table}[t]
\centering
\small
\resizebox{0.98\linewidth}{!}{%
\setlength{\tabcolsep}{3pt}
\begin{tabular}{lccccc}
\hline
\textbf{System} & \textbf{oaoa} & \textbf{oaoo} & \textbf{oaaa} & \textbf{aaaa} & \textbf{cross-profile} \\
\hline
Baseline  & 39.62$\pm$1.19 & 55.42$\pm$3.11 & 60.94$\pm$3.99 & 4.90$\pm$1.52  & 30.89$\pm$1.27 \\
VOXLET    & 27.75$\pm$1.72 & 36.50$\pm$4.53 & 66.79$\pm$3.13 & 5.87$\pm$1.15  & 50.01$\pm$0.02 \\
RASP      & 38.92$\pm$3.07 & 48.82$\pm$7.31 & 88.61$\pm$2.95 & 20.04$\pm$3.53 & 27.57$\pm$2.93 \\
SHADOW    & 45.40$\pm$1.21 & 55.54$\pm$4.42 & 80.49$\pm$3.76 & 5.45$\pm$0.86  & 3.47$\pm$0.57  \\
PHORTRESS & 49.79$\pm$0.60 & 64.08$\pm$3.87 & 86.81$\pm$2.02 & 24.62$\pm$2.08 & 23.28$\pm$3.10 \\
\hline
\end{tabular}
}
\caption{Mean EER (\%) for trials, aggregated across all test sets and SID backends. Values are shown as \emph{mean $\pm$ half-width of 95\% CI}, computed by non-parametric bootstrap (B = 10000) over the per-condition EER values.}
\label{tab:eer_overall}
\end{table}

\subsubsection{Overall EER Landscape}
Aggregating across all test sets and SID backends, the EER landscape varies substantially by system and trial type. For SID attack trials, mean EERs are shown in Table~\ref{tab:eer_overall}. 

PHORTRESS is closest to the 50\% target on \textit{oaoa} trial, indicating the strongest resistance under the primary attack scenario. A consistent pattern appears in the \textit{oaaa} trial: all systems exceed 50\%, often substantially. This elevation suggests anonymization-induced compression of the speaker identity space, making non-target pairs harder to distinguish and inflating EER above chance.

For verification trials (target: 0\% EER), Baseline, VOXLET, and SHADOW achieve low EER on \textit{aaaa} trial types, indicating that their pseudo-speakers consistently sound like themselves across utterances. PHORTRESS shows the highest EER, suggesting that its anonymized segments for the same speaker can sound noticeably different from one another. Cross-profile results show a sharp contrast: SHADOW achieves low EER, meaning its pseudo-identities are readily distinguishable from one another (pseudo-Alice sounds different from pseudo-Bob). VOXLET is near chance (50\%), meaning its pseudo-identities sound so similar that SID backends cannot tell them apart.

\subsubsection{Privacy--Consistency Trade-off}
\begin{figure}[t]
\centering
\includegraphics[width=\linewidth]{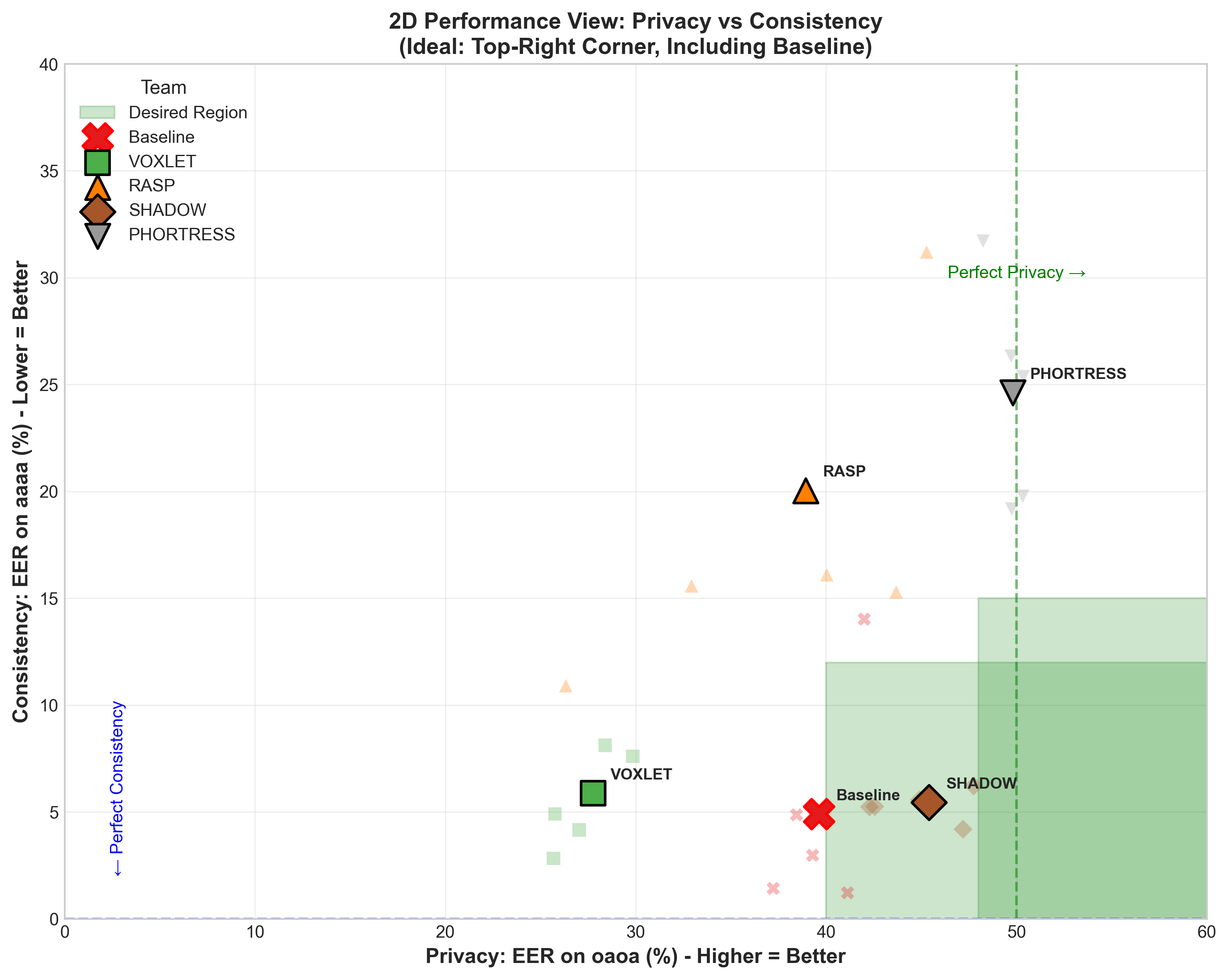}
\caption{Privacy--consistency trade-off across systems. The ideal region combines high EER on \textit{oaoa} trial (better privacy) with low EER on \textit{aaaa} trial (better consistency).}
\label{fig:privacy_consistency_scatter}
\end{figure}

Figure~\ref{fig:privacy_consistency_scatter} illustrates that systems occupy distinct operating points in the privacy vs.\ consistency plane. PHORTRESS is the hardest system to trace back to the original speaker, but its anonymized segments for the same speaker are less reliably recognized as belonging to the same pseudo-identity. SHADOW offers a more balanced profile, maintaining both reasonable privacy and consistent pseudo-speaker output. VOXLET exhibits strong consistency but weaker privacy.

\begin{figure*}[th]
\centering
\includegraphics[width=0.9\linewidth]{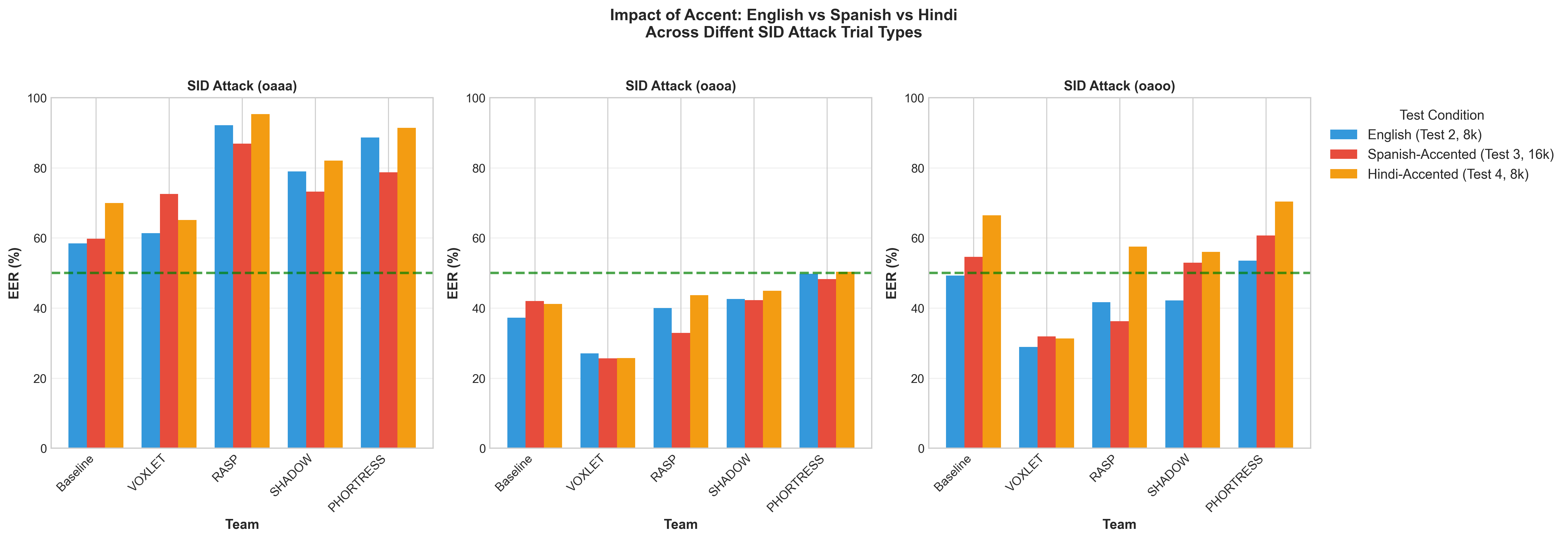}
\hfill
\caption{Impact of speaker accent on SID attack performance. Accent effects on the primary \textit{oaoa} trial are modest, while \textit{oaoo} and \textit{oaaa} show larger system-specific interactions.}
\label{fig:accent_impact}
\end{figure*}

\begin{figure}[t]
\centering
\includegraphics[width=\linewidth]{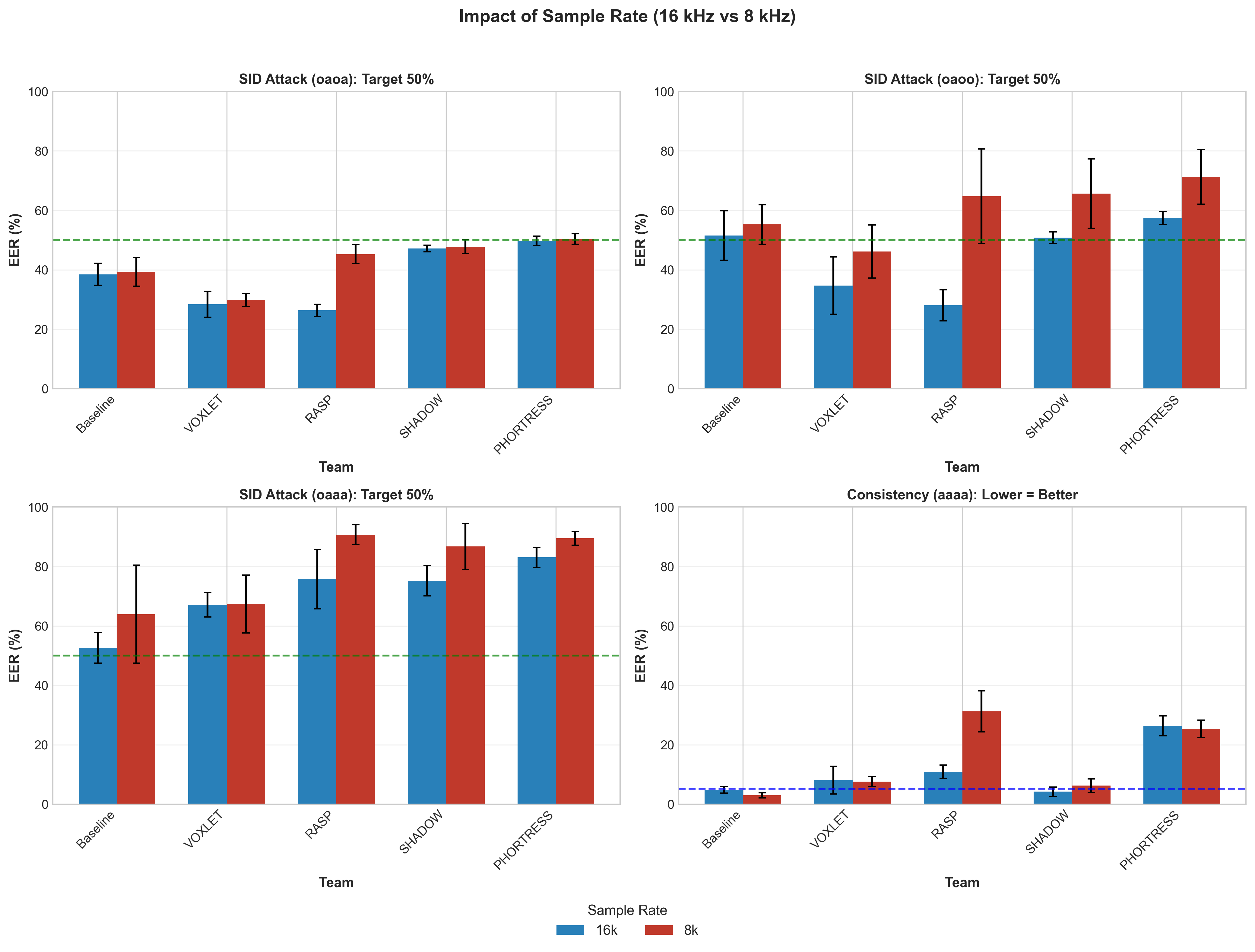}
\caption{Impact of sample rate on speaker de-identification performance for Test 1 dataset (76 English speakers, Mixer 6).}
\label{fig:samplerate_comparison}
\end{figure}

\subsubsection{Accent Effects}
Evaluation on Test 3 and 4 assess robustness to non-native accents. Figure~\ref{fig:accent_impact} shows the results. Differences in \textit{oaoa} trial EER across accent conditions are generally modest relative to inter-team differences. In contrast, the results on \textit{oaoo} and \textit{oaaa} trials exhibit larger shifts for some teams (e.g., Test 4 Hindi increasing \textit{oaoo} EER for every system), suggesting interactions between accent-dependent acoustics and trial-type composition.

\subsubsection{Sample Rate Effects}
Evaluation results on Test 1 dataset shown in Figure~\ref{fig:samplerate_comparison} provide a controlled comparison using the same 76 English speakers at both 16~kHz and 8~kHz. For \textit{oaoa} trial, most systems show stable performance across sample rates, except RASP, which moves from 26.3\% (16~kHz) to 45.3\% (8~kHz), toward the 50\% target. However, RASP’s \textit{aaaa} EER degrades (10.9\% to 31.2\%), indicating that reduced bandwidth affects both attackers and within-system consistency.

\begin{table*}[t]
\centering
\small
\resizebox{\linewidth}{!}{%
\setlength{\tabcolsep}{4pt}
\begin{tabular}{lcccccccl}
\hline
\textbf{System} & \textbf{SBLS} & $P_{\text{attr}}$ & $P_{\text{assoc}}$ & $P_{\text{subgroup}}$ & \textbf{Sex AUC} & \textbf{Age AUC} & \textbf{Max Reident} & \textbf{Most Vulnerable} \\
\hline
Baseline & \textbf{0.877$\pm$0.039} & 0.921$\pm$0.063 & 0.997$\pm$0.005 & 0.552$\pm$0.089 & 0.569$\pm$0.048 & 0.510$\pm$0.044 & 45.1$\pm$8.9\% & Adult\_Male \\
VOXLET & \textbf{0.728$\pm$0.027} & 0.645$\pm$0.034 & 0.903$\pm$0.030 & 0.542$\pm$0.055 & 0.855$\pm$0.033 & 0.482$\pm$0.034 & 46.4$\pm$5.5\% & Adult\_Female \\
RASP & \textbf{0.617$\pm$0.021} & 0.515$\pm$0.037 & 0.895$\pm$0.017 & 0.264$\pm$0.041 & 0.922$\pm$0.016 & 0.564$\pm$0.034 & 73.7$\pm$4.1\% & Adult\_Male \\
SHADOW & \textbf{0.593$\pm$0.031} & 0.505$\pm$0.040 & 0.776$\pm$0.030 & 0.403$\pm$0.080 & 0.955$\pm$0.014 & 0.541$\pm$0.037 & 60.3$\pm$7.8\% & Adult\_Male \\
PHORTRESS & \textbf{0.920$\pm$0.011} & 1.000$\pm$0.012 & 0.997$\pm$0.004 & 0.606$\pm$0.050 & 0.464$\pm$0.039 & 0.496$\pm$0.027 & 39.4$\pm$5.0\% & Adult\_Male \\
\hline
Original & \textbf{0.456$\pm$0.023} & 0.368$\pm$0.039 & 0.676$\pm$0.028 & 0.194$\pm$0.031 & 0.973$\pm$0.008 & 0.659$\pm$0.039 & 81.4$\pm$3.0\% & Adult\_Female \\
\hline
\end{tabular}
}
\caption{SBLS results (sex + age). Higher SBLS indicates better privacy. AUC of 0.5 corresponds to chance-level attribute prediction. Values are shown as \emph{mean $\pm$ half-width of 95\% CI}, computed by a speaker-clustered non-parametric bootstrap (B = 1000).}
\label{tab:sbls_overall}
\end{table*}

\subsection{SBLS Results: Soft Biometric Leakage}

\subsubsection{Overall Results (Sex + Age)}
Table~\ref{tab:sbls_overall} summarizes SBLS using sex and age. All SDID systems improve over the unprocessed baseline, indicating meaningful but incomplete protection. Across systems, $P_{\text{assoc}}$ is consistently high, suggesting errors are not driven by simple, exploitable mappings. The dominant differentiator is subgroup protection $P_{\text{subgroup}}$, where residual vulnerability remains.

Sex is consistently harder to mask than age: several systems yield near-perfect sex recovery (AUC $\approx 0.95$) even when age leakage is moderate. This asymmetry is consistent with deeply embedded sex-correlated acoustic cues (e.g., fundamental frequency and formant structure).

SBLS rankings are stable across test sets (Table~\ref{tab:sbls_per_dataset}). PHORTRESS ranks first and Baseline second in every condition; other systems swap only in isolated cases.

\begin{table}[t]
\centering
\small
\resizebox{\linewidth}{!}{%
\begin{tabular}{lccccc}
\hline
 & \textbf{test1-16k} & \textbf{test1-8k} & \textbf{test2} & \textbf{test3} & \textbf{test4} \\
\hline
\textbf{Baseline} & 0.816$\pm$0.080 & 0.812$\pm$0.086 & 0.863$\pm$0.057 & 0.847$\pm$0.092 & 0.846$\pm$0.108 \\
\textbf{VOXLET} & 0.665$\pm$0.072 & 0.659$\pm$0.074 & 0.724$\pm$0.047 & 0.634$\pm$0.096 & 0.698$\pm$0.065 \\
\textbf{SHADOW} & 0.547$\pm$0.045 & 0.552$\pm$0.065 & 0.542$\pm$0.049 & 0.629$\pm$0.053 & 0.522$\pm$0.100 \\
\textbf{RASP} & 0.605$\pm$0.045 & 0.611$\pm$0.054 & 0.634$\pm$0.046 & 0.566$\pm$0.057 & 0.588$\pm$0.040 \\
\textbf{PHORTRESS} & 0.863$\pm$0.042 & 0.879$\pm$0.037 & 0.908$\pm$0.024 & 0.893$\pm$0.070 & 0.931$\pm$0.024 \\
\hline
\end{tabular}
}
\caption{SBLS with sex and age attributes per dataset. Values are shown as \emph{mean $\pm$ half-width of 95\% CI} from a speaker-clustered non-parametric bootstrap (B = 1000).}
\label{tab:sbls_per_dataset}
\end{table}

\subsubsection{Accent as a Third Attribute}
Adding accent family preserves top and bottom rankings while narrowing mid-tier differences (Table~\ref{tab:sbls_accent}). PHORTRESS remains near-chance on accent (AUC $\approx 0.51$), whereas VOXLET exhibits substantial accent leakage (AUC $\approx 0.78$), producing the largest SBLS drop. RASP and SHADOW improve due to stronger accent subgroup protection.

\begin{table}[t]
\centering
\small
\resizebox{\linewidth}{!}{%
\begin{tabular}{lcccc}
\hline
\textbf{System} & \textbf{SBLS (accent)} & \textbf{SBLS (no accent)} & $\boldsymbol{\Delta}$ & \textbf{Accent AUC} \\
\hline
\textbf{Baseline} & 0.857$\pm$0.029 & 0.877$\pm$0.039 & -0.020$\pm$0.020 & 0.610$\pm$0.034 \\
\textbf{VOXLET} & 0.674$\pm$0.024 & 0.728$\pm$0.027 & -0.054$\pm$0.018 & 0.775$\pm$0.026 \\
\textbf{RASP} & 0.652$\pm$0.021 & 0.617$\pm$0.021 & +0.035$\pm$0.014 & 0.692$\pm$0.026 \\
\textbf{SHADOW} & 0.649$\pm$0.020 & 0.593$\pm$0.031 & +0.056$\pm$0.016 & 0.710$\pm$0.019 \\
\textbf{PHORTRESS} & 0.911$\pm$0.009 & 0.920$\pm$0.011 & -0.009$\pm$0.010 & 0.509$\pm$0.020 \\
\hline
Original & 0.453$\pm$0.022 & 0.456$\pm$0.023 & -0.003$\pm$0.016 & 0.857$\pm$0.020 \\
\hline
\end{tabular}
}
\caption{SBLS with and without accent as a third attribute. Values are shown as \emph{mean $\pm$ half-width of 95\% CI} from a speaker-clustered non-parametric bootstrap (B = 1000); the $\Delta$ CI is paired (accent and no-accent SBLS are evaluated on the same resample each iteration).}
\label{tab:sbls_accent}
\end{table}

\subsubsection{Weight Sensitivity}
\label{sec:weight_sensitivity}

We recompute SBLS under seven weighting schemes that vary 
$(\alpha,\beta,\gamma)$ for attribute predictability, systematic association, and subgroup protection:
default (0.4,0.4,0.2), equal (0.33,0.33,0.34), 
$\alpha$-heavy (0.6,0.2,0.2), 
$\beta$-heavy (0.2,0.6,0.2), 
$\gamma$-heavy (0.2,0.2,0.6), 
no-sub (0.5,0.5,0.0), and 
min-sub (0.45,0.45,0.1).

Results remain stable (Table~\ref{tab:sensitivity_weights}): the top-3 and bottom-1 are invariant, and only RASP and SHADOW swap under equal and $\gamma$-heavy weighting. This reflects SHADOW’s stronger subgroup protection versus RASP’s higher $P_{\text{attr}}$ and $P_{\text{assoc}}$. Removing or downweighting subgroup protection does not alter overall conclusions.

\begin{table}[t]
\centering
\small
\resizebox{0.98\linewidth}{!}{%
\begin{tabular}{lccccccc}
\hline
\textbf{System} & \textbf{default} & \textbf{equal} & $\boldsymbol{\alpha}$-heavy & $\boldsymbol{\beta}$-heavy & $\boldsymbol{\gamma}$-heavy & \textbf{no-sub} & \textbf{min-sub} \\
\hline
Baseline & 2 & 2 & 2 & 2 & 2 & 2 & 2 \\
VOXLET & 3 & 3 & 3 & 3 & 3 & 3 & 3 \\
RASP & 4 & \textbf{5} & 4 & 4 & \textbf{5} & 4 & 4 \\
SHADOW & 5 & \textbf{4} & 5 & 5 & \textbf{4} & 5 & 5 \\
PHORTRESS & 1 & 1 & 1 & 1 & 1 & 1 & 1 \\
\hline
Original & 6 & 6 & 6 & 6 & 6 & 6 & 6 \\
\hline
\end{tabular}
}
\caption{SBLS rankings across weight configurations.}
\label{tab:sensitivity_weights}
\end{table}


\subsubsection{Component Correlation}
With sex and age only, SBLS components are strongly positively correlated (Pearson $r>0.80$), with perfect rank correlation between $P_{\text{attr}}$ and $P_{\text{assoc}}$ (Spearman $\rho=1.0$). Including accent yields weaker correlations for accent subgroup protection: $\rho=0.20$ with $P_{\text{attr}}$, $\rho=0.31$ with $P_{\text{assoc}}$, and $\rho=0.43$ with $P_{\text{subgroup\_noaccent}}$, supporting accent as a distinct privacy dimension.

\subsection{CMC Analysis: Speaker Re-Identification}

Table~\ref{tab:cmc_table} reports CMC re-identification rates (lower is better). PHORTRESS achieves the lowest rates, near the random-permutation baseline. VOXLET exhibits substantially elevated rates across ranks, indicating strong residual identity preservation in embedding space. TitaNet typically yields slightly higher re-identification than ECAPA-TDNN, suggesting increased sensitivity to residual speaker cues.

\begin{table}[t]
\centering
\small
\resizebox{0.98\linewidth}{!}{%
\begin{tabular}{lcccc}
\hline
\textbf{System} & \textbf{k=1 (ECAPA)} & \textbf{k=1 (TitaNet)} & \textbf{k=20 (ECAPA)} & \textbf{k=20 (TitaNet)} \\
\hline
Baseline & 0.043 & 0.055 & 0.228 & 0.263 \\
VOXLET & 0.231 & 0.267 & 0.622 & 0.686 \\
RASP & 0.078 & 0.084 & 0.398 & 0.418 \\
SHADOW & 0.027 & 0.034 & 0.142 & 0.165 \\
PHORTRESS & \textbf{0.016} & \textbf{0.014} & \textbf{0.099} & \textbf{0.101} \\
\hline
\end{tabular}
}
\caption{CMC re-identification rates (lower is better privacy). CMC@k with two NeMo SID backends.}
\label{tab:cmc_table}
\end{table}

\subsection{CCA and Procrustes: Embedding Subspace Analysis}

Table~\ref{tab:cca_proc} summarizes embedding subspace similarity. PHORTRESS is the only system with CCA below the random baseline, indicating stronger-than-random decorrelation between original and anonymized subspaces. All other systems remain well above baseline, suggesting that substantial linear structure is preserved. Procrustes alignment is consistent with CCA trends.

\begin{table}[t]
\centering
\small
\resizebox{0.98\linewidth}{!}{%
\begin{tabular}{lcccc}
\hline
\textbf{System} & \textbf{CCA (ECAPA)} & \textbf{CCA (TitaNet)} & \textbf{Proc (ECAPA)} & \textbf{Proc (TitaNet)} \\
\hline
Baseline & 0.903 & 0.910 & 0.591 & 0.595 \\
VOXLET & 0.900 & 0.915 & 0.597 & 0.624 \\
RASP & 0.877 & 0.899 & 0.581 & 0.617 \\
SHADOW & 0.870 & 0.874 & 0.552 & 0.557 \\
PHORTRESS & \textbf{0.694} & \textbf{0.707} & \textbf{0.390} & \textbf{0.407} \\
\hline
Random & 0.734 & 0.734 & 0.291 & 0.306 \\
\hline
\end{tabular}
}
\caption{CCA of top 10 and Procrustes alignment results averaged across datasets and pseudo-profiles. Lower CCA and Procrustes similarity (Proc) indicate stronger decorrelation.}
\label{tab:cca_proc}
\end{table}

\subsection{Speech Intelligibility: The Privacy--Utility Trade-off}

Table~\ref{tab:overall_utility} reports overall intelligibility aggregated across datasets and ASR configurations. VOXLET achieves the best utility, while PHORTRESS exhibits the largest utility degradation. This highlights a sharp privacy--utility trade-off: the most privacy-protective system is also the least intelligible, whereas the most intelligible system offers weaker privacy on several dimensions.

\begin{table}[t]
\centering
\small
\resizebox{0.98\linewidth}{!}{%
\begin{tabular}{lcc}
\hline
\textbf{System} & \textbf{Average WER} & \textbf{Average Cosine Similarity} \\
\hline
VOXLET & \textbf{0.24} & \textbf{0.860} \\
RASP & 0.44 & 0.686 \\
SHADOW & 0.37 & 0.729 \\
PHORTRESS & 0.70 & 0.482 \\
\hline
\end{tabular}
}
\caption{Overall intelligibility of anonymized segments from SDID systems (lower WER and higher cosine similarity indicate better utility).}
\label{tab:overall_utility}
\end{table}

\begin{table}[t]
\centering
\small
\resizebox{0.98\linewidth}{!}{%
\begin{tabular}{lccccc}
\hline
\textbf{System} & \textbf{Test1-16k} & \textbf{Test1-8k} & \textbf{Test2} & \textbf{Test3} & \textbf{Test4} \\
\hline
VOXLET & 0.26 & 0.23 & 0.22 & 0.27 & 0.24 \\
RASP & 0.24 & 0.51 & 0.47 & 0.29 & 0.61 \\
SHADOW & 0.27 & 0.32 & 0.32 & 0.45 & 0.39 \\
PHORTRESS & 0.57 & 0.59 & 0.62 & 0.82 & 0.75 \\
\hline
\end{tabular}
}
\caption{WER by dataset/accent condition (averaged across ASR models).}
\label{tab:wer_by_dataset}
\end{table}

\section{Discussion}
This study demonstrates that privacy in speaker de-identification is inherently multi-dimensional. Evaluation practices often rely on a single metric—most commonly speaker verification EER—while other signals of identity leakage remain less frequently measured. Our results show that EER, CMC, embedding subspace similarity, and SBLS each probe a distinct mechanism of identity retention: pairwise matching, gallery-based retrieval, global structural preservation, and demographic predictability, respectively. Notably, the weak correlations between certain SBLS components, particularly accent subgroup protection with others, suggest that accent attributes may occupy partially independent representational subspaces.

These findings carry clear methodological implications: reliance on a single metric can mischaracterize privacy risk, since a system may reduce speaker verification accuracy while preserving embedding-level structure or demographic predictability; subgroup-level analyses are essential, as aggregate performance can obscure uneven protection across demographic groups; and privacy must be assessed alongside utility, as identity suppression can degrade linguistic fidelity under distributional shifts such as non-native accents or reduced bandwidth. Notably, WER and semantic similarity capture linguistic content preservation but not perceived naturalness, speaker consistency, signal-level quality, or forensic detectability; human listening studies, speech-quality measures, and audio-forensic detectors would complement the current framework.

Finally, subspace decorrelation below random-pair baselines suggests that some anonymization strategies actively restructure representational geometry rather than merely perturbing it—a distinction future work could formalize using information-theoretic or geometric measures.

\section{Conclusion}
We evaluated speaker de-identification systems using five complementary metrics, showing that they capture different aspects of identity leakage and can yield divergent privacy assessments. These findings motivate multi-metric evaluation as standard practice and point toward anonymization methods that better balance privacy protection, speech utility, and broader measures of speech quality.

\section{Acknowledgements}
This research is based upon work supported by the Office of the Director of National Intelligence (ODNI), Intelligence Advanced Research Projects Activity (IARPA), Anonymous Real-Time Speech (ARTS) research program, under Interagency Agreement (IAA) with NIST IARPA-20001-D250300042. The views and conclusions contained herein are those of the authors and should not be interpreted as necessarily representing the official policies or endorsements, either expressed or implied, of the ODNI, IARPA, NIST or the U.S. Government.



{\small
\bibliographystyle{ieee}
\bibliography{egbib}
}

\end{document}